\documentclass[12pt]{article}
\usepackage[utf8]{inputenc}
\usepackage[english]{babel}
\usepackage{amsmath, amsthm, amssymb, fancyhdr, fancybox, graphicx,bm}
\usepackage{multirow}
\usepackage{pdflscape}
\usepackage{rotating}
\usepackage{booktabs}
\usepackage[authoryear]{natbib}
\usepackage{floatrow}
\usepackage{caption}
\usepackage{authblk}
\usepackage{multirow}
\usepackage{subcaption}
\usepackage{dsfont}
\usepackage{rotating}

\begin{document}
\title{\Large \bf An Imputation model by Dirichlet Process Mixture of Elliptical Copulas for  Data of Mixed Type}
\author[1,2]{\small Jiali Wang\thanks{Corresponding author: Jiali Wang, Data61, Commonwealth Scientific and Industrial Research Organisation, Canberra, ACT 2601, Australia. Email: jiali.wang@data61.csiro.au}}
\author[2]{\small Anton H. Westveld}
\author[2]{\small Bronwyn Loong}
\author[2]{\small Alan H. Welsh}
	
\affil[1]{\scriptsize Data61, Commonwealth Scientific and Industrial Research Organisation, AUS} 
\affil[2]{\scriptsize Research School of Finance, Actuarial Studies and Statistics, College of Business and Economics, The Australian National University, AUS} 

\date{}
\maketitle
\fancyfoot[L]{\today\ \currenttime}

\begin{abstract}
Copula-based methods provide a flexible approach to build missing data imputation models of multivariate data of mixed types.  However, the choice of copula function is an open question. We consider a Bayesian nonparametric approach by using an infinite mixture of elliptical copulas induced by a Dirichlet process mixture to build a flexible copula function. A slice sampling algorithm is used to sample from the infinite dimensional parameter space. We extend the work on prior parallel tempering used in finite mixture models to the Dirichlet process mixture model to overcome the mixing issue in multimodal distributions. Using simulations, we demonstrate that the infinite mixture copula model provides a better overall fit compared to their single component counterparts, and performs better at capturing tail dependence features of the data. Simulations further show that our proposed model achieves more accurate imputation especially for continuous variables and better inferential results in some analytic models. The proposed model is applied to a medical data set of acute stroke patients in Australia. 	
\end{abstract}

\section{Introduction}
Missing data is a common occurrence in multivariate data sets, and the direct application of standard statistical techniques proposed for complete data sets may produce invalid statistical inference. Under the Bayesian framework, missing data is treated as a source of uncertainty, and conditional on the observed data, we impute the missing values as well as make inference on unknown parameters. Multiple Imputation (MI), which was proposed by \cite{rubin1976inference}, takes every $m^{th}$ imputed value from the marginal distribution of the missing values ($m$ needs to be large to ensure independence between consecutive imputed values) to form several `complete' data sets, and the methods designed for complete data sets can be applied to each imputed `complete' data set. Combining rules \citep{rubin1987multiple} are applied to obtain a single inferential result for the quantity of interest. 

In this paper, we consider two major difficulties in building an imputation model that together have not been thoroughly researched yet: a) dealing with heterogeneity in the population and b) modeling variables of mixed-type. Regarding heterogeneity, there have been many proposed imputation models in the literature, but with a few exceptions, they assume the unit records are independent and identically distributed or any grouping structure is known. However, it is not uncommon for a data set to have underlying subpopulations which might not be known a priori. In this situation, mixture models can be used to describe data when potential clusters exist. However selecting the number of clusters is another challenge. To take advantage of the flexibility of nonparametric models, we consider a Dirichlet Process Mixture (DPM) model to induce an infinite mixture model, where the grouping structure is detected by the data and the model complexity is allowed to grow with the sample size. Using finite/infinite mixture models to impute continuous missing values has been considered by \cite{di2007imputation,kim2014multiple}, and to impute unordered categorical missing values \citep{si2013nonparametric,manrique2014bayesian}. 

Another difficulty with building an imputation model is the presence of mixed-type variables (continuous, ordered categorical and unordered categorical). Most of the existing imputation models can only handle one particular type, or a combination of two. In addition, surprisingly very few papers distinguish between ordered categorical data and unordered categorical data, however the former indeed contains extra information \cite[Chapter~12]{hoff2009first}. The most popular method to impute variables of mixed type is the fully conditional specification (FCS), or called the chaining method \citep{van2007multiple}. In this method, every variable with missing values is imputed based on a regression model allowing for different models for each variable, and Gibbs-type loops run through all the regression models to impute missing values until convergence. Though this method is conceptually easy to implement, the major critique of FCS is the lack of theoretical justification to ensure convergence to the proper joint model, and specifying many regression models is labor intensive. 

In this paper, our aim is to develop a flexible imputation model using a Bayesian nonparametric approach. Specifically, an infinite mixture of elliptical copulas induced by a DPM model. Our proposed model accounts for the potential heterogeneity in the population by using a mixture model and variables of mixed-type are handled by combining the extended rank likelihood of the Gaussian copula model \citep{hoff2007extending} for continuous and ordinal variables and a multinomial probit model \citep{albert1993bayesian} for nominal variables. On the computational side, we use the slice sampling algorithm \citep{walker2007sampling} and prior parallel tempering algorithm \citep{van2015overfitting} to perform an exact sampling (no truncation in the infinite mixture model) from a DPM, and also to aid mixing in the MCMC in a multimodal target distribution.  Similar goals have been pursued by other authors, for example, \cite{murray2016multiple} fused two DPM of multivariate Gaussian distributions for continuous variables and multinomial distributions for categorical variables, and added another DPM to capture local dependence between categorical variables and continuous variables similar to a general location model.  However, they did not distinguish between ordinal variables and nominal variables, and they used a version of the truncated Dirichlet Process \citep{ishwaran2001gibbs} which was required to choose the truncation level of the number of components. 

The outline of this paper is as follows: in Section 2 we review the DPM model and extended rank likelihood of elliptical copulas, and describe our proposed DPM of elliptical copulas imputation model to impute mixed-type variables. In Section 3, we focus on the computational aspects of our proposed model, as sampling from Bayesian nonparametric models is challenging and there are mixing issues when sampling from multimodal distributions. Section 4 contains three simulation studies to examine the model fit and assess the imputation accuracy for our proposed approach compared to some alternatives when missing data are involved. A simulation study based on a real clinical data set is also presented. Section 5 concludes the article and discusses some items for future research.

\section{Model Specification}
\subsection{Dirichlet Process Mixture}
We start by introducing some fundamental concepts of a Dirichlet Process (DP) that are relevant to this article (for a more comprehensive review, refer to Chapter 1 and 2 in \cite{muller2015bayesian}). 

\subsubsection{Construction of a Dirichlet Process}
A nonparametric model is characterized by an infinite number of parameters. Suppose we have a collection of data $y_1,...,y_N$, which follow a distribution $G$. The goal is to make inference about $G$ which is in an infinite dimensional space. To proceed with a Bayesian nonparametric model, we need a prior on $G$ known as a Bayesian nonparametric prior. Among a range of Bayesian nonparametric priors, we will focus on the DP prior because of its mathematical convenience. A DP($MG_0$) is characterized by two quantities, the total mass parameter $M$ and the centering measurement $G_0$. For each partition $\{B_1,...,B_K\} $ on a set $B$, DP($MG_0$) assigns probability $G(B_k)$ to every subset $B_k$, such that $G(B_1),...,G(B_K) \sim Dir(MG_0(B_1),...,MG_0(B_K))$. From the definition we can see clearly the analogy between a DP and a Dirichlet distribution: a DP is an infinite dimensional extension to a Dirichlet distribution. In addition, the mathematical convenience of a DP comes from its conjugacy. Let $y_1,...,y_N|G \sim G$, and let $G \sim DP(MG_0)$, then the posterior also follows a DP: $G|y_1,...,y_N \sim DP(MG_0+\sum_{i=1}^N \delta_{y_i}) = DP\big((M+N)(\frac{M}{M+N}G_0+\frac{1}{M+N}\sum_{i=1}^N\delta_{y_i})\big)$,where the centering measurement is a weighted average of the prior distribution and point masses, and the total mass parameter increases by $N$.

 A DP can be constructed by several equivalent ways, e.g., Chinese restaurant process, Polya urn process, and the stick breaking construction. In this paper we will use the stick breaking construction because it will facilitate computation. Specifically, by the stick breaking construction, $G$ can be represented as an infinite sum of point masses with different weights $w_h$ and locations $\theta_h$, such that $G(\cdot)=\sum_{h=1}^{\infty}w_h \delta_{\theta_h}(\cdot)$, where $\delta(\cdot)$ is the Dirac function, $\theta_h \sim G_0, v_h \sim Beta(1,M)$ and $w_h=v_h \prod_{l < h}(1-v_l)$. In this way, a DP induces clusters, as data within a cluster share the same parameter $\theta_h$. In a DP, the centering measurement $G_0$ is the expectation of the distribution $G$ before data are collected, therefore it represents the prior belief on $G$. The total mass parameter $M$ controls how concentrated the induced distribution is around $G_0$, as a bigger $M$ generates more clusters and a distribution closer to $G_0$. To illustrate this, in Figure 1, we generated three DPs by the stick breaking construction, with $G_0=N(0,1)$ and $M=1,10,100$ respectively. When $M=1$, only a few clusters were generated with bigger weights and the induced distribution was far apart from the standard normal distribution. On the  other hand, when $M=100$, a greater number of clusters were generated with smaller weights and the induced distribution was very close to the standard normal distribution. We will make use of this fact in the prior tempering algorithm to sample from a DPM model in the next section.

\begin{figure}
	\center
	\includegraphics[width=10cm, height=6cm]{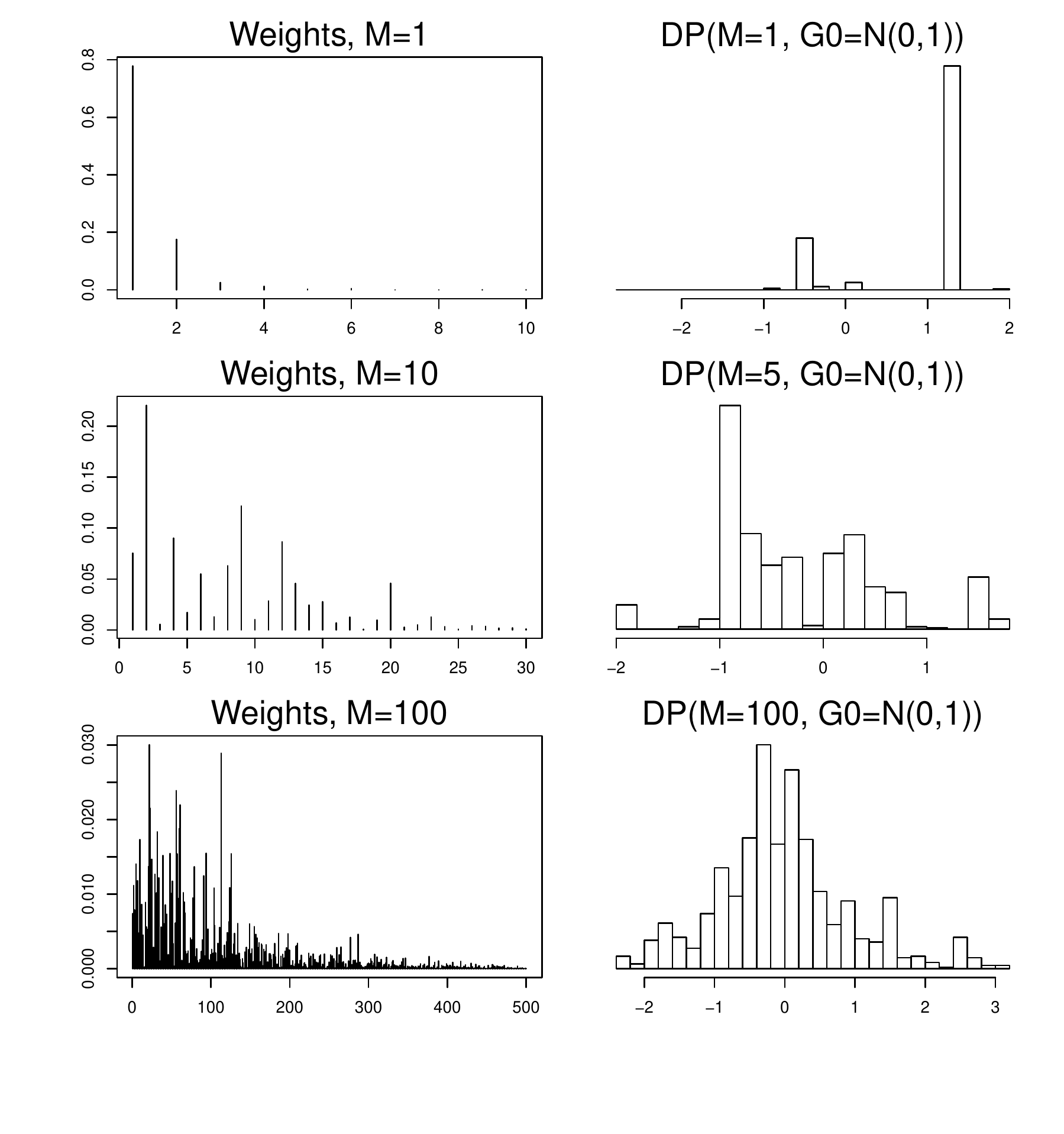}
	\caption{Dirichlet Processes generated by stick breaking construction, with $G_0=N(0,1), M=1,10,100$ respectively.}
\end{figure}

\subsubsection{Dirichlet Process Mixture}
From the stick breaking construction we can see that a DP only generates discrete distributions, and it is not directly generalizable to continuous distributions. To overcome this limitation, DPM models can be used by putting a DP on the distribution of the parameters in a parametric kernel: 
\begin{equation}
f_G(y)=\int f_{\theta}(y)dG(\theta), G \sim DP(MG_0)
\end{equation}
Equivalently,  a DPM can be represented hierarchically, where each data point $y_i$ is associated with its own parameter $\theta_i$ and the distribution of those parameters follow a DP:
\begin{equation}
y_i|\theta_i \sim f_{\theta_i}, \theta_i|G \sim G, G \sim DP(MG_0)
\end{equation}
In this representation, clusters are induced by the grouping of parameters. A third equivalent representation is constructed through the stick breaking of a DP which induces an infinite mixture model:
\begin{equation}
f(y|w,\theta)=\sum_{h=1}^{\infty}w_hf(y|\theta_h)
\end{equation}
It shares many of the same interpretations and properties as a finite mixture model \citep{mclachlan2004finite}, but is more flexible in terms of allowing the model complexity to be adaptable from data. 

\subsubsection{Selecting the Number of Components in a Finite Mixture Model}
In a finite mixture model $f(y_i|w,\theta)=\sum_{h=1}^H w_hf(y_i|\theta_h)$, the number of clusters $H$ is fixed although unknown, and some model selection procedures should be applied to choose a `best' $H$. This is known as the order selection problem \cite[Chapter~6]{mclachlan2004finite}. Some commonly used methods for the order selection problem involve calculating an information criteria (AIC, BIC), performing a LRT hypothesis test, and calculating the Bayes factor between two competing models, etc. These methods require that a number of potential models are fit that one is chosen among them. As such, these methods can be quite computationally intensive and may fail because of nonidentifiability issue and use of improper priors. Specifically, the nonidentifiability issue arises in the mixture distribution when the model can be represented equivalently well by different $H$, and this may lead to an unbounded likelihood function in a frequentist model or a divergent posterior if an improper prior is specified in a Bayesian model. In addition, the methods also fail to account for the variability in $H$ if it is treated as fixed. Some fully Bayesian procedures consider model selection and parameter estimation in a single MCMC run by treating $H$ as a random variable. For example, the reversible jump algorithm \citep{richardson1997bayesian} allows the chain to jump between different $H$ values corresponding to different models until convergence; in a birth and death process \citep{stephens2000bayesian} the parameters in the model are viewed as a point process and new components are allowed to be `born' and existing components to `die'; and by prior parallel tempering \citep{van2015overfitting}, where far more clusters than supported by the data are initially included and the redundant clusters are gradually removed or merged with other clusters in the MCMC by controlling priors. We have chosen to use the prior parallel tempering algorithm in our proposed model because it neither requires to design a delicate balance criteria in trans-dimensional algorithms, nor extra programming effort is needed in the parallel chains by just changing the hyperparameters that resemble temperatures. Furthermore, the prior parallel tempering algorithm achieves order selection and assists mixing simultaneously.

\subsection{Extended Rank Likelihood of Elliptical Copulas}
In this section we introduce the elliptical copula models as the mixing distribution in the DPM model. Copula models are often used to analyse variables of different types in multivariate data sets. A copula model describes the relationships among variables by decomposing the joint distribution function of variables $Y_1,...,Y_p$ into the marginal distributions $F_1(Y_1),...,F_p(Y_p)$, and a copula function $C$, as implied by Sklar's theorem: $F(Y_1,...,Y_p)=C(F_1(Y_1),...,F_p(Y_p))$.

\subsubsection{Inference of a Copula Model}
Inference of a copula model involves estimating the marginal distribution functions $F_l$ and the copula function $C$. Estimating the marginal distribution functions can be done either parametrically or non-parametrically. A more challenging task is the choice of a copula function which is no easier than specifying a multivariate distribution function directly. Nevertheless, copula models are still advantageous to use, especially when there are mixed-type data involved, because it breaks down a complex task into two simpler tasks - variables are first transformed to the same uniform scale from the marginal distribution specifications, and then their associations are modeled on the latent variables space. Choosing a copula function is usually based on the knowledge of the data, for example by plotting the data to see if a pair of variables exerts left and/or right tail dependence, and if extreme values exist. Another strategy is to construct a copula function from a mixture of existing families, as the convex combination of copula functions is still a proper copula function. \cite{hu2006dependence} studied the dependence patterns across financial markets by constructing a mixture of bivariate Gaussian copula, Gumbel copula, and its survival copula. \cite{manner2011tails} provided some insights into the tail dependence of a mixture of elliptical copulas, and demonstrated its use in a time series stock market data set. \cite{wu2014bayesian,wu2015bayesian} investigated the DPM of (skew) Gaussian copulas of continuous variables only.

\cite{hoff2007extending} proposed a semi-parametric approach to estimate the copula function by the rank of the data without explicitly specifying the marginal distributions, when the goal is to understand the relationships among variables, and it provides a unified framework to model ordered variables. As in \cite{hoff2007extending} we focus on the elliptical class of copulas, including the Gaussian copula and t copula. Specifically, let $u_j=F_j(y_j), j=1,...,p$, then the Gaussian copula and t copula are:

\begin{equation}
\begin{aligned}
&C_G(u_1,...,u_p|\Sigma)=\Phi_p(\Phi^{-1}(u_1),...,\Phi^{-1}(u_p)|\Sigma),\\
&C_t(u_1,...,u_p|\Sigma,\nu)=\Phi_p(t^{-1}(u_1),...,t^{-1}(u_p)|\Sigma,\nu),
\end{aligned}
\end{equation}
where $\Phi/\Phi_p, t/t_p$ are respectively the univariate/p-variate distribution functions of the Gaussian distribution and the t distribution with degrees of freedom equal to $\nu$. 

In the Gaussian copula, let $z_{ij}=\Phi^{-1}(F_j(y_{ij}))=\Phi^{-1}(u_{ij})$, then the latent variables $z$ follow a Gaussian distribution with the correlation matrix $\Sigma$: $z_1,...,z_N|\Sigma \sim N_p(0,\Sigma)$. Because we know $\Phi^{-1}(F_j(\cdot))$ is a non-decreasing monotone transformation, the order of the latent variables $z$ is consistent with the order of the observed data $y$, provided that there is a meaningful ordering of data (continuous variables and ordinal variables). Let $D(y)$ denotes the set of all possible $z$ which are consistent with the ordering of $y$. In other words, those $z$ satisfy $z_{ij} < z_{tj}$ if $y_{ij} < y_{tj}$, and $z_{ij} > z_{tj}$ if $y_{ij} > y_{tj}$. \cite{hoff2007extending} claimed that it is partial sufficient to make inference on the correlation matrix $\Sigma$ based on the extended rank likelihood function $p(z \in D|\Sigma)$. 

Computationally, we sample the latent variables $z$ from multivariate normal distributions with truncations, to meet the constraint of the ordering of the observed data. Specifically, for each unit $i$ in the $j^{th}$ variable, the lower bound is the maximum value in the $j^{th}$ latent variable $z_j$ whose corresponding $y$ is smaller than $y_{ij}$. The upper bound can be defined accordingly. That is, the lower bound (lb) and upper bound (ub) of $z_{ij}$ are defined as: $lb=max\{z_{tj}: y_{tj} < y_{ij}\},ub=min\{z_{tj}: y_{tj} > y_{ij}\}$. Having sampled the latent variables $z$, we estimate the correlation matrix $\Sigma$, which is expected to inherit the dependency structure in the observed data $y$. 

Similarly for t copulas, we just replace the normal distribution functions with the t distribution functions and the additional degrees of freedom parameter $\nu$ is needed. The idea of the extended rank likelihood of Gaussian copulas can be paralleled to the t copulas without too much modifications. In this paper, we consider both the Gaussian copula and t copula, however, we illustrate our model specification and sampling algorithm for the Gaussian copula, because of the nice conditional Gaussianity and the natural link to the multinomial probit model for unordered data. In addition, we will show in the later simulation section that when using a mixture model, the difference between using a Gaussian kernel and t kernel is small in terms of overall fitting and preservation of some key quantities.

\subsection{Dirichlet Process Mixture of Elliptical Copulas}
In this section, we layout the specification of our proposed model. We combine the extended rank likelihood and the multinomial probit model for variables with and without ordering respectively. 

Suppose there are $p+1$ random variables in total, and without loss of generality assume the first $p$ variables are continuous or ordinal, and the $(p+1)^{th}$ variable is nominal with $Q+1$ categories (for example 0,1,...,Q). In order to avoid cluttered notations we only include one nominal variable, however, extension to several nominal variables is straightforward. The DPM is applied on the latent variable $\mathbf{z_i}=(\mathbf{z_{i,1:p}},\mathbf{z_{i,(p+1):(p+Q)}})=(\mathbf{z_{i,(1)}},\mathbf{z_{i,(2)}}),i=1,...,N$, where the vector $\mathbf{z_{i,(1)}}$ is the latent variable of length $p$ for the first $p$ variables of observation $i$, and the vector $\mathbf{z_{i,(2)}}$ is the latent variable of length $Q$ for the nominal variable. Specifically, for the nominal variable, category $q~(q=1,...,Q)$ is observed if the $q^{th}$ element in the vector $\mathbf{z_{i,(2)}}$ is the largest and is greater than 0, otherwise  the baseline category ($q=0$) is observed if all the elements in the vector $\mathbf{z_{i,(2)}}$ are smaller than 0. According to the stick breaking representation, the DPM of a Gaussian copula with the extended rank likelihood and the multinomial probit model is:

\begin{equation}\label{main model}
\begin{aligned}
&\mathbf{z_i}|\bm{\beta},\Sigma, w=\sum_{h=1}^{\infty} w_h N_{p+Q}(\mathbf{z_i}|(\mathbf{0},\bm{\beta_h}),\Sigma_h), \\
&w_h=v_h \prod_{l<h}(1-v_l), \text{where} ~v_l \sim Beta(1,M),\\
&y_{ij}=F_j^{-1}(\Phi(z_{ij})), j=1,...,p,\\
&y_{i,p+1}=\left\{
\begin{array}{l}
q, \text{if max$(\mathbf{z_{i,(2)}}) >0$ and is the $q^{th}$ element in $\mathbf{z_{i,(2)}}$} \\
0, \text{if max$(\mathbf{z_{i,(2)}})<0$}
\end{array}
\right.
\end{aligned}
\end{equation}

In model (\ref{main model}), the latent variable $z_i$ follows a mixture of multivariate Gaussian distributions, with the weight $w_h$ for cluster $h$, and $\sum_{h=1}^{\infty}w_h=1$. The mean for $z_i$ is $(\mathbf{0},\bm{\beta_h})$, where $\mathbf{0}$ is a vector of length $p$ for the ordered variables, and $\bm{\beta_h}$ is a vector of length $Q$ representing the relative difference between each of the categories in the nominal variable compared to the baseline category.  The correlation matrix $\Sigma_h$ is identifiable by fixing the variances along the diagonal to be 1. It can be split into blocks such that $\Sigma_h=\begin{pmatrix}
	\Sigma_{h,1} & \Sigma_{h,12} \\
	\Sigma_{h,21} &  \Sigma_{h,2}
\end{pmatrix}$,
where $\Sigma_{h,1}$ is the correlation matrix of the latent variables corresponding to the ordinal variables, $\Sigma_{h,2}$ is the correlation matrix of the latent variables for each categories of $Y_{p+1}$, and $\Sigma_{h,12}(\Sigma_{h,21})$ is the association between the two sets of variables. 

We assume the centering measurement $G_0$ in the DP for the means and correlations are independent, such that $G_0(\bm{\beta},\Sigma)=G_0(\bm{\beta})G_0(\Sigma)$. We further assume $G_0(\bm{\beta}) \sim N(\mu_{\beta}, \Lambda_{\beta})$, and this is equivalent to putting a semi-conjugate prior on $\bm{\beta_h}$. However, as there is no semi-conjugate prior on a correlation matrix, we follow the procedure in \cite{hoff2007extending} to work with an expanded model instead by putting an Inverse Wishart prior on the variance covariance matrix $\tilde \Sigma_h$: $G_0(\tilde \Sigma) \sim Inv-Wishart (\nu_{\Sigma}, \Lambda_{\Sigma})$ and then rescale them to correlation matrix $\Sigma_h$ in each iteration in the MCMC. To incorporate the weakest prior information, the hyperparameters in the simulation studies were chosen to be $\bm{\mu_{\beta}}=\mathbf{0}$, $\Lambda_{\beta}=I_Q$, $\nu_{\Sigma}=Q+2$, $\Lambda_{\Sigma}=I_{p+Q}$, where $I_d$ stands for the identity matrix of dimension $d$. 

\section{Computation}
Sampling from DPM models is a well known difficult task because the random process contains infinite dimensional parameters and the multimodal surface may hinder Markov chain achieving irreducibility condition. We start with an overview of some popular sampling algorithms of DPM models and some methods to aid mixing when sampling from multimodal distributions, and then describe our proposed sampling method. 

\subsection{Sampling from Dirichlet Process Mixture Models}
 Sampling from DPM models has long been dominated by integrating out the random measurement $G$ \citep{escobar1994estimating}. This approach relies on the Polya urn scheme to sample the parameters for each observation conditional on the rest, that is, from the distribution $p(\theta_i|\theta_{-i},y)$, thus creating a Markov chain. Although many variations have been proposed to speed up mixing in the chain, this method is still argued to suffer from slow convergence because of the high dependency among parameters. \cite{ishwaran2001gibbs} proposed to truncate the infinite summation at a certain level by bounding the approximation error. Intuitively it is applicable because the infinitesimal weights will not play a key role, however it is not an exact sampling algorithm. \cite{walker2007sampling} and \cite{papaspiliopoulos2008retrospective} independently developed exact sampling algorithms called a slice sampler and retrospective sampler for a DPM model. The difference between these two algorithms lies in the allocation of observations to clusters. The slice sampler introduces latent variables to make the summation finite in an augmented model and allocates observations to those finite clustes with probabilities proportional to their likelihood. On the contrary, the retrospective sampler proposes the allocation of observations to clusters through a Metropolis Hastings step but the acceptance balance criteria is difficult to construct. Next we will briefly review the key steps of constructing a slice sampler, and will leave the details of the implementation specific to our proposed model to Section 3.3. 
 
 Recall the stick breaking representation of a DPM model is
 $f(y_i|w,\theta)= \sum_{h=1}^{\infty} w_h f(y_i|\theta_h)$. The essential
 step to eliminate the infinite summation in the slice sampler is the introduction of latent variables $u_i$ for each of the observations. That is, $p(y_i,u_i|w,\theta)=\sum_{h=1}^{\infty}\mathds{1}_{(u_i<w_h)} f(y_i|\theta_h)$, such that there are only a finite $k^*$ number of terms in the summation, where $k^*$ is the minimum integer to satisfy $\sum_{h=1}^{k^*}w_h >1-u^*, u^*=min(u_1,...,u_n)$. The next step is to introduce the group indicators $r_i$ to each observation to identify which group the observation belongs to, then the augmented model becomes $p(y_i,u_i,r_i|w,\theta)=\mathds{1}_{(u_i<w_h)}f(y_i|\theta_{r_i})$. From here we are able to construct a Gibbs sampler for the augmented model to update $w_h$ (through $v_h$), $\theta_h$, $u_i$ and $r_i$ iteratively \cite[Chapter~2]{walker2007sampling,muller2015bayesian}. Notice that in the Gibbs sampler, we generate the cluster-specific parameters $w_h$ and $\theta_h$ from the existing clusters from the previous iteration, and if extra clusters are needed ($k^* >max (r_i)$), then we sample $w_h$ (through $v_h$) and $\theta_h$ from their prior distributions.

\subsection{Mixing Issues and Label Switching}
 Another potential issue with sampling from DPM models stems from poor mixing of the Markov chain. When the posterior distribution is multimodal with well separated modes, MCMC methods are prone to being trapped in local regions and cannot explore the whole parameter space. A phenomenon due to implementing a Bayesian simulation based estimation procedure of a mixture model is called label switching. Under noninformative priors on the clusters, if we interchange the parameters as well as the weights in any two clusters, the likelihood remains invariant, therefore the sampler cannot uniquely identify the cluster labels in the MCMC. Theoretically, there will be $H$ factorial modes in the marginal posterior distribution in each component if there are $H$ distinct clusters and if the sampler mixes well. The presence of label switching indicates good mixing, but at the same time the symmetry of the marginal distributions means it is not appropriate to simply calculate the posterior summaries of component-specific parameters. Some previous work has been done to solve the label switching problem for the purpose of posterior summaries in finite mixture models. For example, \cite{richardson1997bayesian} imposed order constraints on the parameters to break the symmetry; \cite{hurn2003estimating} proposed a label invariant loss function approach to select the permutation which minimises the loss. Handling the label switching problem for an infinite mixture model is more challenging. \cite{van2015overfitting} proposed a `Zswitch' algorithm for a finite mixture model by relabeling the samples. However, in our paper the presence of label switching will not affect the inferential result, because our focus is on posterior density estimation to build a flexible imputation model when heterogeneous subpopulations might exist, rather than untangling cluster-specific parameters. In fact, adequate label switchings indicates a good mixing.  
 
 To illustrate a situation of poor mixing with no label switching, we generated bivariate data from a two-component distributions: $\frac{1}{2}N((-1,3),I_2)+\frac{1}{2}N((2,1),I_2)$. The likelihood has two well separated modes, and we implemented the slice sampler algorithm to estimate the mean parameters in the normal distributions. In this simple example we may argue that the sampler was estimating the true parameter values as shown in the traceplot of the MCMC, but did not achieve good mixing because the sampler fails to get out of local regions (see Figure 2(a)). 
  
 \begin{figure}
 	\begin{subfigure}[b]{0.4\textwidth}
 		\includegraphics[width=6.3cm,height=5.3cm]{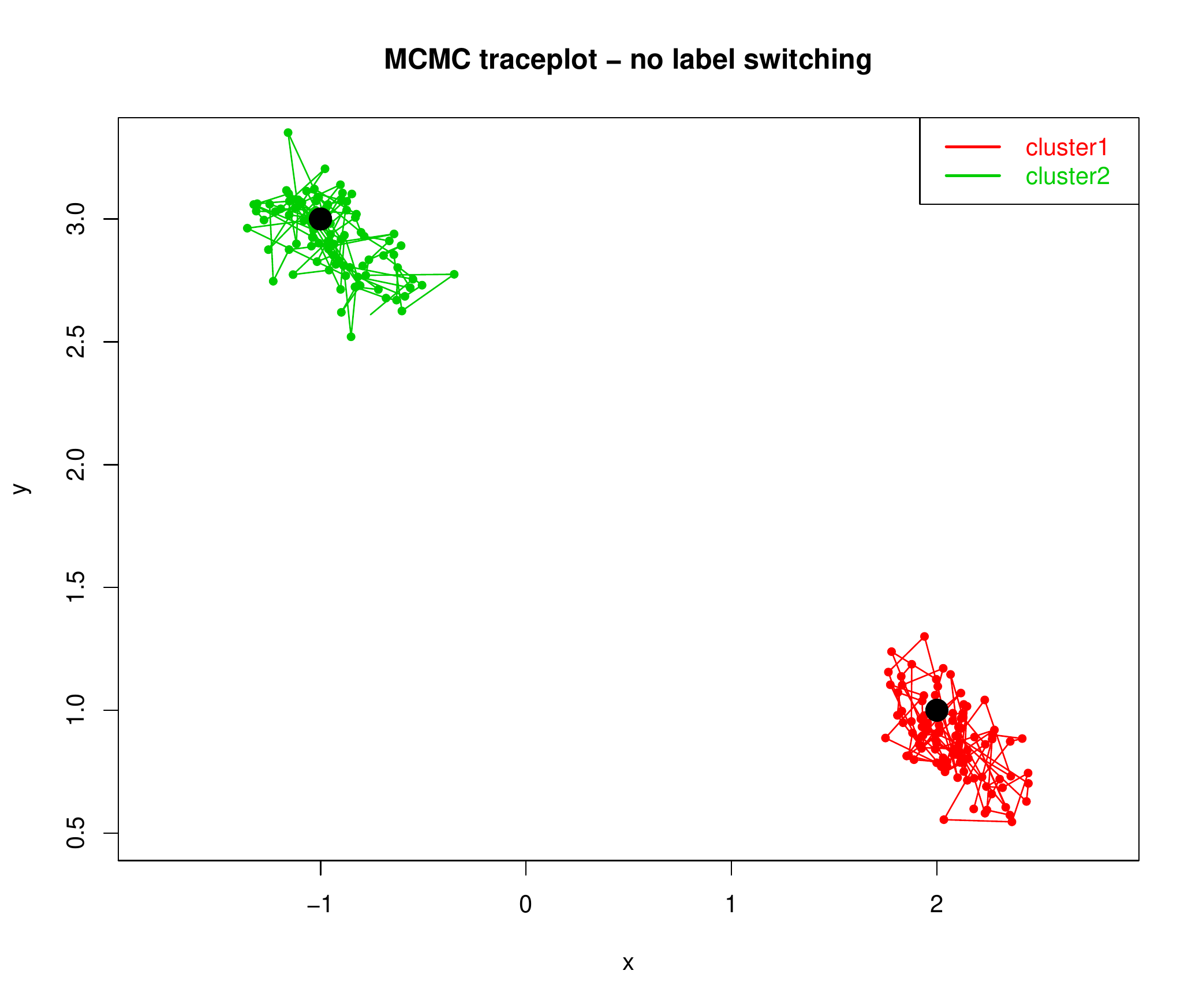}
 		\label{datanolabelswitching}
 	    \subcaption{}
 	\end{subfigure}
 	\begin{subfigure}[b]{0.4\textwidth}
 		\includegraphics[width=6cm,height=5cm]{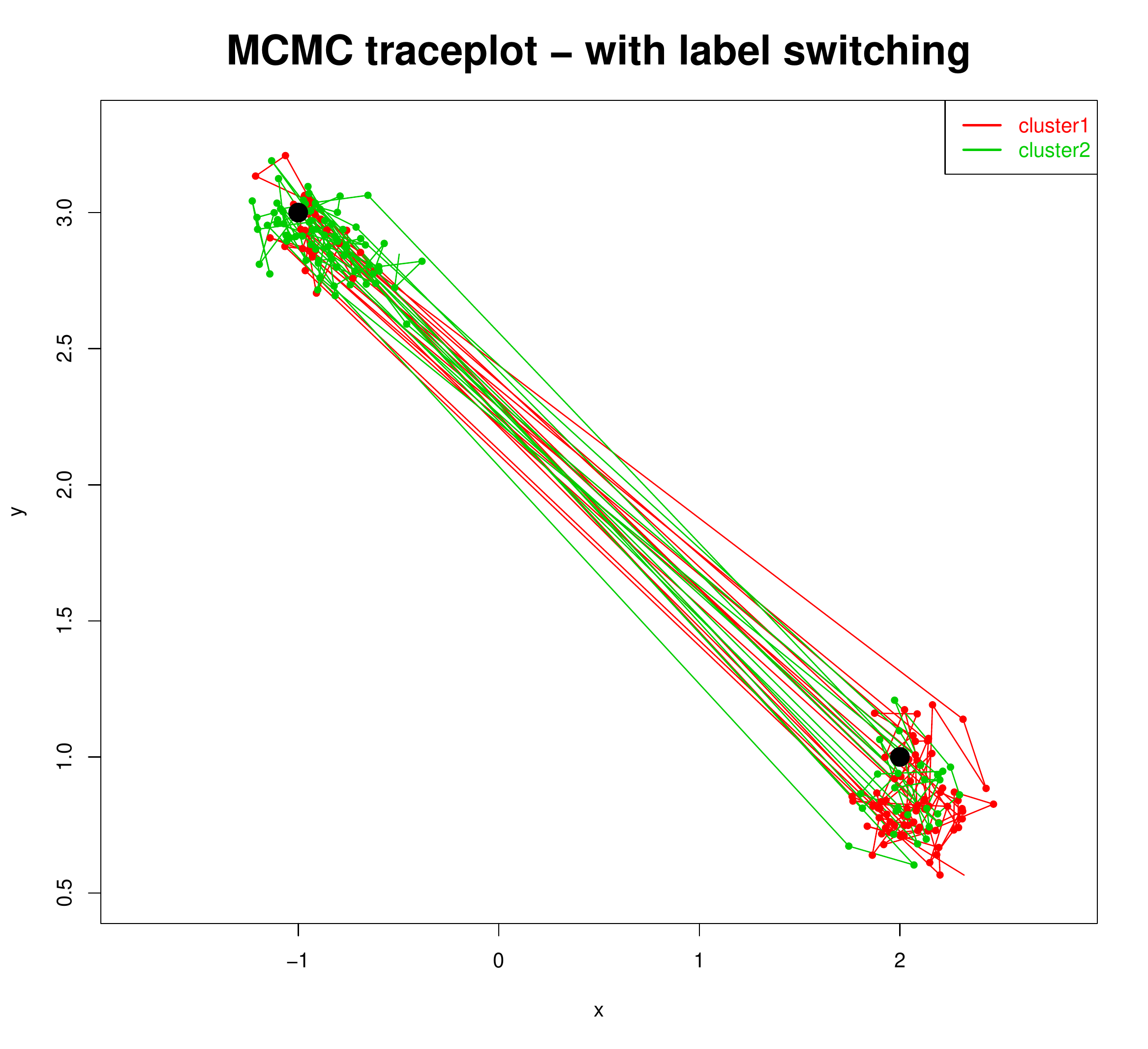}
 		\label{nolabelswitching}
 		\subcaption{}
 	\end{subfigure}
 \caption{Data generated from a two component normal mixture distribution: $\frac{1}{2}N((-1,3),I_2)+\frac{1}{2}N((2,1),I_2)$. The figures show trace plots of the mean parameters in a MCMC sampler when (a) we implement a slice sampling algorithm of DPM alone, no label switching between two modes; (b) we implement a slice sampling algorithm with prior parallel tempering with 4 chains, label switching occurs between the two modes.}
 \end{figure}

 \subsection{Prior Parallel Tempering}
 There are some proposed solutions in the literature to help the sampler move more freely, for example, instead of updating parameters through a Gibbs sampler, a Metropolis Hasting step can be used. \cite{papaspiliopoulos2008retrospective} and \cite{hastie2015sampling} adopted a label switching move through a Metropolis Hasting step in the MCMC but it is difficult to construct a good acceptance probability of switching. Another approach, parallel tempering, originated from physics and was introduced to statistics by \cite{neal1996sampling}. The idea is to flatten the target distribution by raising it by power to certain levels to facilitate movement between modes, while the locations of modes in the target distribution do not change. Specifically, define $p^{\alpha}$, where $p$ is the target distribution we want to sample from and $\alpha$ acts as the tempering parameter. With a higher temperature ($\alpha \to 0$) , the density $p^{\alpha}$ is more flattened. Higher temperatures allow a sampler to move more freely, while lower temperatures can reflect the target distribution more precisely. Therefore, it is a trade off between precision and efficiency. To run the parallel tempering algorithm, we set up several chains with gradual increasing temperatures in a sequence, then we run the MCMC in each chain with its own temperature respectively, and before moving to the next iteration, we swap the parameters in two adjacent chains (two chains with consecutive temperatures) with some balancing probabilities, which will be shown shortly. \cite{van2015overfitting} applied the parallel tempering idea in the Bayesian estimation of the finite mixture model by choosing different hyperparameters in the Dirichlet prior on weights, and they referred to it as the prior parallel tempering algorithm.
 
 Combining the slice sampling algorithm and the prior parallel tempering algorithm, we outline our proposed algorithm specific to the DPM model. Our algorithm achieves order selection and aids mixing simultaneously by choosing different hyperparameters of total mass parameter $M$ in the DP prior. As discussed in Section 2.1.1, $M$ controls the concentration of the distribution, with a smaller $M$ inducing fewer cluster with bigger weights for each cluster, and a bigger $M$ leading to better mixing but may induce redundant clusters. 
 
 \begin{itemize}
 	\item Step 1: Set up $K$ chains with increasing total mass parameters ($M_1,...,M_K$); 
 	\item Step 2: For each iteration $s=1,...,S$ in the MCMC, run the Gibbs sampler separately in each chain, draw $\theta_{[h,M_K]}^{(s)}, w_{[h,M_K]}^{(s)} (v_{[h,M_K]}^{(s)}), u_{[i,M_K]}^{(s)}$ and $r_{[i,M_K]}^{(s)}$, $h=1,...,H^{(s-1)}, i=1,...,N$. If extra clusters are needed, draw $\theta_{[h,M_K]}^{(s)}, w_[{h,M_K]}^{(s)} (v_{[h,M_K]}^{(s)}), h=H^{(s-1)}+1, ..., H^{(s)}$ from the priors;
 	\item Step 3: Before going to the next iteration, swap everything except for the total mass parameters in the adjacent chains $k$ and $k'=k+1$ with probability: min\big(1,$\frac{G_k(\theta_{M_{k'}}|Y)G_{k'}(\theta_{M_{k}}|Y)}{G_k(\theta_{M_{k}}|Y)G_{k'}(\theta_{M_{k'}}|Y)}$\big), where $G_k(\theta_{M_k}|Y)$ is the posterior distribution of $\theta$ in the $k^{th}$ chain. Calculating this acceptance ratio is made possible by the link of a DP to the Dirichlet distribution applied to a finite data set. 
 	\item Repeat step 2 and 3 until convergence.
  \end{itemize}

After running the prior parallel tempering algorithm, we choose the chain with the lowest temperature to be our target chain to perform posterior inference, because it reflects more accurate local estimation and more parsimonious clustering of data. 
 
For a simple illustration of our proposed algorithm, we added the prior parallel tempering to the slice sampling algorithm on the same data set in Section 3.2,  by setting up 4 chains ($M=0.005,0.01,0.05,0.1$) in the DPM model with Gaussian kernels. This time, we can see that the sampler was switching between the two modes frequently and still obtained accurate estimates of the correct parameters (Figure 2(b)). We can further confirm that the label switching indeed occurred by keeping track of the swapping among the four chains in the MCMC. 

\subsection{Algorithm of the DPM of Gaussian Copulas with Missing Values of Mixed-type}
In this section, we outline the steps to implement the DPM of Gaussian copula with missing values of mixed-type. The sampling model and the priors are specified in Section 2.3. For each scan $s=1,...,S$:

$\triangleright$ For each $M_k$ in $(M_1,...,M_K)$:\\
1. Estimate the parameters in each of the $K$ parallel chains separately:
\begin{itemize}
	\item Sample $\beta_{[h,M_k]}, h=1,...,H_{M_k}$\\
	$\beta_{h,M_k} \sim N_Q (\mu_{[h,M_k]}, V_{[h,M_k]} )$, \\
	where $A_{[h,M_k]}=\#\{i:r_{i,M_k}=h\}$, is the number of observations in group $h$, ~$\mu_{[h,M_k]}= (\Lambda_{\beta}^{-1} + A_{[h,M_k]} V_{[h,M_k]}^{-1})^{-1}(\Lambda_{\beta}^{-1} \mu_{\beta}+ V_{[h,M_k]}^{-1}(z_{[(2),r_{i,M_k}=h]}-z_{[(1),r_{i,M_k}=h]}\Sigma_{[h,21,M_k]}\Sigma_{[h,1,M_k]}^{-1}))$,~
	$V_{[h,M_k]}=\frac{1}{A_{[h,M_k]}} (\Sigma_{[h2,M_k]} - \Sigma_{[h,21,M_k]})\Sigma_{h,1,M_k}^{-1}\Sigma_{[h,12,M_k]}$
	
	\item Sample $\tilde \Sigma_{[h,M_k]}, h=1,...,H_{M_k}$\\
	$\tilde \Sigma_{h,M_k} \sim \text{Inverse Wishart} (\nu_{\Sigma}+A_{h,M_k}, (\Lambda_{\Sigma}+\epsilon_{[r_{i,M_k}=h]}^T\epsilon_{[r_{i,M_k}=h]}))$\\
	where $\epsilon_{[r_{i,M_k}=h]}=z_{[r_{i,M_k}=h]}-(0,\beta_{[h,M_k]})$\\
	Rescale to correlations matrices $\Sigma_{[h,M_k]}[a,b]= \tilde \Sigma_{[h,M_k]}[a,b] /\sqrt{\tilde \Sigma_{[h,M_k]}[a,a] \tilde \Sigma_{[h,M_k]}[b,b]}$
	
	\item Sample $w_{[h,M_k]}$ (through  $v_{[h,M_k]}, h=1,...,H_{[M_k]}$)\\
	$v_{[h,M_k]}\sim Beta(1+A_{[h,M_k]}, M_k+B_{[h,M_k]})$, where $\#B_{[h,M_k]}=\{i:r_{[i,M_k]}>h\}$, 
	$w_{[h,M_k]}=v_{[h,M_k]} \prod_{l<h}(1-v_{[l,M_k]})$
	
	\item Sample $u_{[i,M_k]},i=1,...,N$\\
	$u_{[i,M_k]} \sim \text{Unif}(0,w_{[r_{i,M_k}]})$
	
	\item Sample $r_{[i,M_k]},i=1,...,N$ from the multinomial distribution\\
	$p(r_{[i,M_k]}=h)\propto L(z_{[i,M_k]}|\beta_{[r_i,M_k]},\Sigma_{[r_i,M_k]}) \mathds{1}_{(w_{[h,M_k]}>u_{[i,M_k]})}$\\
\end{itemize}

2. Impute missing values in ordinal/continuous variables ($y_1,...,y_p$) by the extended rank likelihood

\hspace{0.1mm}For each $i$ in $1:N$ and for each $j$ in $1:p$
\begin{itemize}
	\item Compute $lb=max(z_{[tj,M_k]}:y_{tj}<y_{ij})$, and $ub=min(z_{[tj,M_k]}:y_{tj}>y_{ij})$\\
	Sample $z_{[ij,M_k]}\sim N\big((\Sigma_{[r_{i,M_k]}}[j,-j]\Sigma_{[r_{i,M_k}]}[-j,-j]^{-1})(z_{[i,-j,M_k]}-(0_{-j},\beta_{[r_i,M_k]}),\Sigma_{[r_{i,M_k}]}[j,j]-\Sigma_{[r_{i,M_k}]}[j,-j]\Sigma_{[r_{i,M_k}]}[-j,-j]^{-1}\Sigma_{[r_{i,M_k}]}[-j,j]\big) \mathds{1}_{(lb,ub)}$, which is a truncated normal distribution for non-missing $y_{ij}$, and a normal distribution without the truncation for missing $y_{ij}$;
	\item If $y_{ij}$ is missing, then $y_{[ij,M_k]}^{mis}=\hat F_j^{-1}(\Phi(z_{[ij,M_k]}))$, where $\hat F_j$ is the empirical CDF of variable $j$.\\
\end{itemize}
3. Impute missing values in nominal variables by the multinomial probit model
\begin{itemize}
\item Sample $z_{[i,(2),M_k]} \sim N\big(\beta_{[r_i,M_k]}+\Sigma_{[21,r_{i,M_k}]}\Sigma_{[1,r_{i,M_k}]}^{-1}z_{[i,(1),M_k]}, \Sigma_{[2,r_{i,M_k}]}- \Sigma_{[21,r_{i,M_k}]}\Sigma_{[1,r_{i,M_k}]}^{-1}\Sigma_{[12,r_{i,M_k}]}\big)$
\item Accept the draw until the mapping rule in the model specification (\ref{main model}) is satisfied if $y_{i,p+1}$ is missing, and always accept the draw if $y_{i,p+1}$ is not missing;
\item If $y_{i,p+1}$ if missing, impute the category by the mapping rule as in (\ref{main model}).
\end{itemize}

$\triangleright$ Exchange chain $k$ and $k'=k+1$ with probability: min(1,$\frac{G_k(\theta_{M_{k'}}|Y)G_{k'}(\theta_{M_{k}}|Y)}{G_k(\theta_{M_{k}}|Y)G_{k'}(\theta_{M_{k'}}|Y)}$). In words, exchange $w_{M_k}$ and $w_{M_k'}$, $\beta_{M_k}$ and $\beta_{M_k'}$, $\Sigma_{M_k}$ and $\Sigma_{M_k^{'}}$, $u_{M_k}$ and $u_{M_k'}$, $r_{M_k}$ and $r_{M_k^{'}}$.

\section{Simulation Experiments}
We now investigate the performance of our proposed model against some competing alternatives by simulation experiments. We ran 3 simulations studies: (i) Simulate bivariate data from two-cluster Gaussian copulas and t copulas, and assess the adequacy of model fit; (ii) Simulate 4-variate latent variables from two-cluster Gaussian copulas and use CDF transformations to generate mixed-type variables. Then we create artificial missingness assuming MAR and compare the accuracy of imputed values; (iii) Simulations based on a real clinical data set. 

\subsection{Simulation 1}
We first begin with a discussion on tail dependence measures as their measures form part of our assessment criteria for simulation. While rank-based measurements of dependency are same for all the elliptical copulas, such as Kendall's $\tau$ and Spearsman's $\rho$, there are some fundamental differences between Gaussian copulas and t copulas, for example, tail dependency. The coefficient of tail dependence summarizes the pair-wise dependence at extremes, and it measures the concurrence of two events. The coefficient of upper and lower penultimate tail dependence $\lambda_u(u)$ and $\lambda_l(u)$ is defined accordingly as follows:

\begin{equation}\label{penultimate}
\begin{aligned}
&\lambda_u(u)=p(F_1(X)>u|F_2(Y)>u)\\
&\lambda_l(u)=p(F_1(X)<u|F_2(Y)<u)
\end{aligned}
\end{equation}
where $u$ is a quantile value. Then the upper and lower coefficients of tail dependence ($\lambda_u$ and $\lambda_l$) are obtained by taking limit in (\ref{penultimate}):

\begin{equation}\label{tail dependence}
\begin{aligned}
&\lambda_u=lim_{u \to 1^-}\lambda_u(u)\\
&\lambda_l=lim_{u \to 0^+}\lambda_l(u)
\end{aligned}
\end{equation}
In the case of elliptical copulas, they display symmetric tail dependence, namely $\lambda_u=\lambda_l=\lambda$. However, the Gaussian copula has asymptotic independence in the tails as $\lambda=0$, whereas t copulas exert positive tail dependence such that $\lambda=2t_{\nu+1}(-\sqrt{\nu+1}\sqrt{\frac{1-\rho}{1+\rho}})$, where $\nu$ is the degrees of freedom and $\rho$ is the correlation coefficient in the scale matrix. Nevertheless t copulas offer a more robust way to analyze data, and as the degrees of freedom increases the t copula becomes indistinguishable from the Gaussian copula. In addtion, when constructing new copulas from mixture models, Manner and Segers (2009) showed that a correlation mixture of elliptical copulas have stronger (penultimate) tail dependence, for the same unconditional correlation of single copulas. This implies that although a mixture of Gaussian copulas is asymptotically independent in the limit, at the penultimate level there is stronger dependence, therefore it falls into the category of near asymptotic dependence.

\subsubsection{Data Generation}
(i) We generated 100 data sets of bivariate variables with $N=200$ units, from a two-component mixture of Gaussian copulas:

\begin{equation*}
\mathbf{U} \sim 0.75 C_G(\mathbf{u}|\Sigma_1) + 0.25 C_G(\mathbf{u}|\Sigma_2),
\end{equation*}
where $\Sigma_1=\begin{pmatrix}
1&-0.6\\
-0.6 &1
\end{pmatrix}$ and 
$\Sigma_2=\begin{pmatrix}
1&0.8\\
0.8 &1
\end{pmatrix}$.
\\
(ii) Similarly, we generated 100 data sets of bivariate variables with $N=200$ units, from a two-component mixture of t copulas, with the same correlation matrices in (i), and the additional degrees of freedom parameters equal to 2 and 4 respectively:
\begin{equation*}
\mathbf{U} \sim 0.75 C_t(\mathbf{u}|\Sigma_1,\nu_1=2) + 0.25 C_G(\mathbf{u}|\Sigma_2,\nu_2=4).
\end{equation*}

\subsubsection{Assessment Criteria}
(i) \underline{LPML}. Firstly we want to assess the adequacy of overall model fit. One approach is to compute the $log~ pseudo~ marginal~ likelihood$ (LPML) (\cite{geisser1979predictive}), which is the Bayesian version of `leave-one-out' cross validation. 
\begin{equation}\label{LPML}
\begin{aligned}
\text{LPML} &= log (\prod_{i=1}^N p(y_i|y_{-i})) = \sum_{i=1}^N log(p(y_i|y_{-i})), \\
p(y_i|y_{-i}) &=\int p(y_i|\theta,y_{-i})) p(\theta|y_{-i})) d\theta \approx (\frac{1}{L}\sum_{l=1}^L \frac{1}{p(y_i|\theta_l)})^{-1},
\end{aligned}
\end{equation}
where $\theta_l$ is the $l^{th}$ draw from the posterior distribution $p(\theta|y)$. LPML is the log of the product of $N$ marginal distributions $p(y_i|y_{-i})$. The product is called the conditional predictive ordinate (CPO). However, the CPO seldom has a closed form expression, therefore, a simulation based approach - harmonic mean of parameters drawing from the posterior distribution - is often used instead. A larger LPML value indicates a better overall model fit. 
\\
(ii) \underline{Posterior Predictive Check}. We also generate multiple replicated data sets from the posterior predictive distribution, and then compared how well the replicated data sets captured some key statistics of the given data set. To do this, we generated multiple replicated data sets from $p(y^{rep}|y)= \int f(y^{rep}|\theta)p(\theta|y)d\theta$. In the case of no exact expression for $p(y^{rep}|y)$, the same MC technique as generating CPO could be applied. We defined some test statistics $T(y)$  and evaluated the Bayesian p-values as $p\big(T(y^{rep})<T(y)\big)$ and coverage as p\big($T(y) \in 95\%$ credible interval of $T(y^{rep})\big)$. For the purpose of this article, we consider the penultimate tail dependence at quantile levels $u=0.95, 0.9, 0.85$ as the test statistics.  A Bayesian p-value closer to 0.5 and coverage closer to 95\% provide evidence that the model captures the test statistics well. 

\begin{table}[H]\label{sim1.G}
	\centering
	\scalebox{0.8}{
	\begin{tabular}{llllll}
		\hline
		\multicolumn{2}{l}{}    & Single Gaussian Copula & Single t Copula & DPM Gaussian Copulas & DPM t Copulas \\ [0.5ex]\hline\hline
		\multirow{2}{*}{u=0.95} & coverage & 51  & 94  & 92  & 93 \\
		& p-value  & 0.74 & 0.59 & 0.59 & 0.6 \\
		\hline
		\multirow{2}{*}{u=0.9}  & coverage & 50 & 94 & 93  & 93\\
		& p-value  & 0.72  & 0.57 & 0.56 & 0.56\\
		\hline
		\multirow{2}{*}{u=0.85} & coverage & 65 & 94& 94 & 95 \\
		& p-value  & 0.66 & 0.51 & 0.51 & 0.5 \\
		[0.5ex]\hline\hline		
		\multirow{2}{*}{LPML} & mean & 5.005 & 17.228  & 169.635  & 161.756  \\
		& sd  & 4.452  & 5.948  & 17.281 & 24.912\\ \cline{1-6} 
	\end{tabular}}
	\caption{Coverages and Bayesian p-values of tail dependence at u=0.95, 0.9, 0.85 and LPML of the four competing methods - single Gaussian Copula, single t copula, DPM Gaussian copulas and DPM t copulas, when the true model is a mixture of two Gaussian copulas.}
\end{table}

\begin{table}[H]\label{sim1.t}
	\centering
	\scalebox{0.8}{
	\begin{tabular}{llllll}
		\hline
		\multicolumn{2}{l}{} & Single Gaussian Copula& Single t Copula & DPM Gaussian Copulas & DPM t Copulas \\ [0.5ex]\hline\hline
		\multirow{2}{*}{u=0.95} & coverage & 37   & 86 & 93 & 92\\
		& p-value  & 0.81& 0.62 & 0.58 & 0.54 \\
		\multirow{2}{*}{u=0.9}  & coverage & 31 & 91 & 95 & 94\\
		& p-value  & 0.81 & 0.59 & 0.54 & 0.53 \\
		\multirow{2}{*}{u=0.85} & coverage & 38 & 88& 97 & 94\\
		& p-value  & 0.77& 0.55 & 0.48 & 0.51 \\		 
		[0.5ex]\hline\hline
		\multirow{2}{*}{LPML}  & mean & 4.692& 40.100 & 246.228 & 241.227\\
		& sd & 5.295& 10.716 & 24.041& 27.644\\ \cline{1-6} 
	\end{tabular}}
	\caption{Coverages and Bayesian p-values of tail dependence at u=0.95, 0.9, 0.85 and LPML of the four competing methods - single Gaussian Copula, single t copula, DPM Gaussian copulas and DPM t copulas, when the true model is a mixture of two t copulas.}
\end{table}

\subsubsection{Simulation 1 - Results}
The four competing methods that we consider in this simulation are: (i) Single Gaussian copula; (ii) Single t copula; (iii) DPM of Gaussian copulas; (iv) DPM of t copulas.

Table 1 summaries the assessment criteria for the four competing methods when the data were generated from a mixture of two-component Gaussian copulas. The total mass parameters in the tempering step were chosen to be $M=(0.005,0.01,0.05,0.1,0.5,0.8,1.1,1.4,1.7,2)$ to ensure adequate switchings between adjacent chains and that the posteriors were flattened enough with higher temperatures. Choosing these settings for $M$ induced an average number of clusters from 2 with the lowest temperature and gradually increased to an average of 10 with the highest temperature. We ran the MCMC for 10,000 iterations, and discarded the first 5000 iterations as the burn-in samples. The single Gaussian copula method suffered from under coverage of the penultimate tail dependence at all u=0.95, 0.9, 0.85 levels, with Bayesian p values away from 0.5. The other three methods performed reasonably well in capturing the penultimate tail dependencies. It is interesting to see that the single t copula achieved the nominal coverage although the data were generated from a mixture model with both positive and negative correlations. This may be because the tail dependence only describes the two dimensional data in the top right and bottom left corners, rather than an overall goodness of fit measure, and may also be because a t copula with negative correlation can exert positive tail dependence, especially when the degrees of freedom is small. On the other hand, LPML takes every single observation into account to give an overall measurement. For LPML, the single Gaussian copula model performed the worst followed by the single t copula, but results were similar between the DPM of Gaussian copulas and the DPM of t copulas. 

Similarly in Table 2, we summarize the model fit of data generated from a two-component mixture of t copulas. While the single Gaussian copula still suffered from under coverage of penultimate tail dependence, in this scenario the single t copula encountered the same problem. In other words, a single t copula cannot fully describe the tail dependence when the true model is a mixture of two t copulas. For LPML, the DPM of Gaussian copulas and the DPM of t copulas both had larger numbers compared with their single component counterparts, indicating better overall fit. We can conclude from simulation 1 that using mixture models achieves better performance in terms of global and local measurements when subpopulations exist, however, it is not clear whether the DPM of t copulas outperforms the DPM of Gaussian copulas or the other way round. 

\subsection{Simulation 2}
\subsubsection{Data Generation}
In this simulation, we aim to examine the imputation accuracy and the ability to recover parameters in some models of interest. We generated 100 data sets with $N=200$ observations each from a two-component mixture of Gaussian copulas of four variables. The first three variables were ordered, specifically $y_1 \sim Pois (1),~y_2 \sim Gamma(1,3), ~y_3 \sim t(df=2,\mu=2)$, and the fourth variable was a nominal variable with 4 categories so that 3 latent variables were needed. Firstly the 6-variate latent variables $z$ were generated by a mixture of two Gaussian copulas model with equal weights, randomly generated correlation matrices $\Sigma_1$ and $\Sigma_2$, and the parameters $\beta_1$ and $\beta_2$ for the ordinal variable.

 \begin{equation*}
 \begin{aligned}
\mathbf{Z} &\sim 0.5 C_G(\mathbf{z}|\bm{\beta_1},\Sigma_1) + 0.5 C_G(\mathbf{z}|\bm{\beta_2},\Sigma_2),\\
\text{where}&~ \bm{\beta_1}=(0.5,1,-0.5),\bm{\beta_2}=(-1,-0.5,1),\\
\Sigma_1&=\begin{pmatrix}
1 &-0.286& -0.409 &-0.038 &-0.410& -0.305\\
-0.286 & 1&  0.085 & 0.193  &0.665 &-0.588\\
-0.409 & 0.085&  1& -0.378& -0.006&  0.034\\
-0.038 & 0.193 &-0.378 & 1& 0.675 & 0.311\\
-0.410  &0.665& -0.006  &0.675  &1  &0.151\\
-0.305 &-0.588  &0.034  &0.311  &0.151  &1
\end{pmatrix},\\
\text{and}~\Sigma_2&=\begin{pmatrix}
 1& -0.404 & 0.285 & 0.074 &-0.058 & 0.075\\
 -0.404&  1&  0.085& -0.147 & 0.306 & 0.832\\
 0.285 & 0.085 & 1 &-0.501&  0.733 &-0.029\\
  0.074& -0.147 &-0.501&  1& -0.037 &-0.072\\
 -0.058 & 0.306 & 0.733 &-0.037&  1 & 0.061\\
 0.075 & 0.832& -0.029& -0.072 & 0.061  &1
 \end{pmatrix}.
 \end{aligned}
\end{equation*}
Then the first three variables in $\mathbf{Z}$ were transformed to the data $Y$ scale by $y_j=\hat F_j^{-1} (\Phi (z_j))$, and for the nominal variable $Y_4$ the category of each unit was selected to be the index of the largest element in the vector of the last three variables in $z$, and to be 0 if all the three elements were less than 0.  

To introduce missingness, assuming MAR we kept $Y_3$ completely observed, and the probabilities of missing in $Y_1, Y_2$ and $Y_4$ depended on $Y_3$ by the logit link, $p_{mis}=\text{logit} (\gamma*y_3)$. We chose $\gamma$ values to control the levels of missingness in each variable from low (10\%), median (20\%) to high(30\%), corresponding to $\gamma=-1.35, -0.65 ~\text{and}~ -0.31$ respectively. Therefore, for each of the 100 complete data sets, three associated data sets with  different amounts of missingness were created. 

\subsubsection{Assessment Criteria}
(i) \underline{Imputation accuracy}\\
We assessed the imputation accuracy by calculating the average discrepancy between the imputed values and the true values in the multiple `complete' data sets. The discrepancy was measured by the Euclidean distance for the continuous variable $Y_2$, and by the misclassification rate for the discrete variables $Y_1$ and $Y_4$.
\\
(ii)  \underline{Parameter estimates in some models of interest}\\
We are also interested in the inferential results in some analysis models. Rubin's combing rules were applied to obtain point estimates for $\beta_0$,...,$\beta_5$ using the multiply imputed data sets. As an illustration in this simulation, a model of interest was chosen to be a Poisson regression:

\begin{equation}\label{Poisson}
log(E(Y_1))=\beta_0+\beta_1Y_2+\beta_2Y_3+\beta_3Y_{4,1}+\beta_4Y_{4,2}+\beta_5Y_{4,3}
\end{equation}

\rotatebox{90}{\parbox{1\textheight}{
		\scalebox{0.72}{
			\begin{tabular}{|l|llllllllllllllllll}
				\hline
				\multirow{3}{*}{} & \multicolumn{6}{l|}{10\%}                                                   & \multicolumn{6}{l|}{20\%}                                                   & \multicolumn{6}{l|}{30\%}                                                   \\ \cline{2-19} 
				&\multicolumn{2}{l}{Y1} & \multicolumn{2}{l}{Y3} & \multicolumn{2}{l|}{Y4} & \multicolumn{2}{l}{Y1} &\multicolumn{2}{l}{Y3} & \multicolumn{2}{l|}{Y4} & \multicolumn{2}{l}{Y1} & \multicolumn{2}{l}{Y3} & \multicolumn{2}{l|}{Y4} \\ \cline{2-19} 
				& mean       & sd         & mean        & sd        & mean       & \multicolumn{1}{l|}{sd}         & mean       & sd         & mean        & sd        & mean       & \multicolumn{1}{l|}{sd}         & mean       & sd         & mean        & sd        & mean       & \multicolumn{1}{l|}{sd}         \\	\hline
				FCS                & 2.167      & 0.606      & 17.247      & 8.602     & 0.573      & \multicolumn{1}{l|}{0.089}      & 3.025      & 1.031      & 17.385      & 4.611     & 0.580      & \multicolumn{1}{l|}{0.077}      & 3.714      & 1.595      & 18.222      & 4.945     & 0.590      & \multicolumn{1}{l|}{0.079}     \\
				Single copula        & 1.916      & 0.424      & 17.198      & 7.416     & 0.555      & \multicolumn{1}{l|}{0.057}      & 1.913      & 0.511      & 17.854      & 6.032     & 0.556      & \multicolumn{1}{l|}{0.077}      & 2.068      & 0.812      & 17.644      & 4.912     & 0.553      & \multicolumn{1}{l|}{0.066}      \\ 
            	DPM copulas               & 1.884      & 0.263      & 14.393      & 5.143     & 0.539      & \multicolumn{1}{l|}{0.053}      & 1.924      & 0.324      & 15.436      & 4.402     & 0.548      & \multicolumn{1}{l|}{0.066}    &2.043      & 0.438      & 16.386      & 4.077     & 0.554      & \multicolumn{1}{l|}{0.098}      \\ 
				\hline
			\end{tabular}}
			\captionof{table}{Imputation accuracy in the simulated data of mixed type, measured by the average Euclidean distance between the imputed values and the true values for the contious variable $Y2$ and the average misclassification rate for the discrete variable $Y_1$ and $Y_4$, of the three competing methods - FCS, single Gaussian coupla and DPM Gaussian copuals.}
			\label{table_exp_settings}
		}}
		\hspace{1.5cm}
		\rotatebox{90}{\parbox{1\textheight}{
				\scalebox{0.58}{
					\begin{tabular}{|l|llllllllllllllllllllllll}
						\hline
						\multirow{3}{*}{} & \multicolumn{6}{l|}{CC}                                                           & \multicolumn{6}{l|}{FCS}                                                           & \multicolumn{6}{l|}{Single copula}    & \multicolumn{6}{l|}{DPM copulas}                                                          \\ \cline{2-25} 
						& \multicolumn{2}{l}{10\%} & \multicolumn{2}{l}{20\%} & \multicolumn{2}{l|}{30\%} & \multicolumn{2}{l}{10\%} & \multicolumn{2}{l}{20\%} & \multicolumn{2}{l|}{30\%} & \multicolumn{2}{l}{10\%} & \multicolumn{2}{l}{20\%} & \multicolumn{2}{l|}{30\%} & \multicolumn{2}{l}{10\%} & \multicolumn{2}{l}{20\%} & \multicolumn{2}{l|}{30\%} \\ \cline{2-25} 
						& bias       & coverage     & bias       & coverage     & bias       & \multicolumn{1}{l|}{coverage}     & bias       & coverage     & bias       & coverage     & bias       & \multicolumn{1}{l|}{coverage}     & bias       & coverage     & bias       & coverage     & bias       & \multicolumn{1}{l|}{coverage}     & bias       & coverage     & bias       & coverage     & bias       & \multicolumn{1}{l|}{coverage}     \\ \cline{1-1}\hline
						$\beta_0$          & 0.114      & 88           & 0.697      & 89           & 1.286      & \multicolumn{1}{l|}{87}           & 0.376      & 80           & 0.758      & 71           & 1.385      & \multicolumn{1}{l|}{83}           & 0.143      & 92           & 0.704      & 87           & 1.11       & \multicolumn{1}{l|}{92}           & 0.165      & 90           & 0.685      & 91           & 1.067      & \multicolumn{1}{l|}{95}           \\ 
						$\beta_1$          & 0.260      & 94           & 1.134      & 82           & 1.590      & \multicolumn{1}{l|}{84}           & 0.649      & 90           & 1.122      & 91           & 1.610      & \multicolumn{1}{l|}{82}           & 0.13       & 100          & 1.108      & 99           & 1.402      & \multicolumn{1}{l|}{95}           & 0.116      & 100          & 1.011      & 100          & 1.308      & \multicolumn{1}{l|}{97}          \\
						$\beta_2$          & 0.030      & 96           & 0.103      & 92           & 0.316      & \multicolumn{1}{l|}{90}           & 0.039      & 81           & 0.183      & 79           & 0.442      & \multicolumn{1}{l|}{77}           & 0.029      & 94           & 0.095      & 86           & 0.319      & \multicolumn{1}{l|}{86}           & 0.038      & 90           & 0.087      & 87           & 0.301      & \multicolumn{1}{l|}{86}           \\ 
						$\beta_3$          & 0.489      & 96           & 1.274      & 89           & 1.115      & \multicolumn{1}{l|}{89}           & 0.679      & 91           & 1.075      & 87           & 1.088      & \multicolumn{1}{l|}{86}           & 0.624      & 89           & 0.998      & 90           & 0.901      & \multicolumn{1}{l|}{97}           & 0.598      & 93           & 0.953      & 92           & 0.918      & \multicolumn{1}{l|}{96}           \\ 
						$\beta_4$          & 0.127      & 99           & 0.809      & 92           & 1.351      & \multicolumn{1}{l|}{85}           & 0.172      & 84           & 0.833      & 71           & 1.178      & \multicolumn{1}{l|}{79}           & 0.135      & 95           & 0.723      & 87           & 1.384      & \multicolumn{1}{l|}{80}           & 0.174      & 96           & 0.620      & 92           & 1.036      & \multicolumn{1}{l|}{85}           \\ 
						$\beta_5$          & 0.112      & 89           & 0.726      & 83           & 1.320      & \multicolumn{1}{l|}{76}           & 0.167      & 85           & 0.793      & 79           & 1.809      & \multicolumn{1}{l|}{78}           & 0.107      & 91           & 0.646      & 88           & 1.129      & \multicolumn{1}{l|}{91}           & 0.110      & 87           & 0.642      & 87           & 1.148      & \multicolumn{1}{l|}{88}           \\ \cline{1-1}\hline
					\end{tabular}}
					\captionof{table}{A comparison of squared bias of the coefficients estimates and the corresponding 95\% coverage in the simulated data sets under four treatments of missing data - complete case analysis, FCS, single Gaussian copula and DPM Gaussian copulas.}
					\label{table_exp_settings}
				}}

\subsubsection{Simulation 2 - Results}
For each simulation, we ran the Gibbs sampler with 10000 iterations, and abandoned the first half of the iterations as the burn-in period. The 10 total mass parameters were chosen to be $M=(1,1.5,...,5.5)$. We created 10 complete data sets for the multiple imputation procedure.

The summary of imputation accuracy is shown in Table 3. The `single copula' method refers to fitting a single Gaussian copula model with no mixtures and `FCS' refers to the fully conditional specification using the `mi' package in R \citep{su2011multiple} where for each of the variables with missing values, a GLM was specified, and then we iterated among those GLMs to impute missing values until convergence. The FCS method almost always imputed missing values which had the largest Euclidean distance and mis-classfication rates compared to the true values. The DPM model achieved more accurate imputation than the single copula model for the continuous variable $Y_3$, but the two copula methods performed similarly for the discrete variables $Y_1$ and $Y_4$.

Table 4 shows the mean squared bias of the coefficient estimates in the Poisson model (\ref{Poisson}) and the corresponding coverage rates. The `true' parameters were estimated from the complete data sets before the artificial missingness was created, and then over the 100 data sets the mean of the squared bias between the `true' parameters and the MI estimates was calculated for the four competing methods. The 95\% confidence intervals were obtained by calculating the point estimates of the parameters plus and minus 1.96 times the corresponding sampling standard errors. The proportion of the 100 confidence intervals then created which contained the `true' parameter values is the coverage rate. When the missingness was low- 10\%, the complete case analysis performed the best, with the smallest squared bias for most of the parameters. As the missingness rate increased to 20\% and 30\%, the two copula methods outperformed the complete case analysis and FCS. The proposed DPM model achieved slightly lower squared bias and had coverages closer to the nominal 95\% than the single copula model, although the difference was not that pronounced. 

\subsection{Simulation 3}
Lastly, we designed a simulation based on a real data set - the Quality in Acute Stroke Care (QASC) study \citep{middleton2011implementation}. The QASC study was a randomized control trial conducted in New South Wales, Australia. There were 19 acute stroke units in two cohorts participated; among them 10 were assigned to the intervention group and 9 were assigned to the control group. The eligible number of participants was 1696. The researchers were primarily interested in four outcome variables: (1) modified Rankin Scale (an ordinal variable ranging from 0 to 6, measuring the degree of disability or dependence in daily activities); (2) Barthel index (an ordinal variable ranging from 0 to 100, which also measures performance in activities of daily living. It is usually reported as a dichotomized variable with 60 or more or 95 or more as cut points); (3) mean SF-36 mental component summary score; (4) mean SF-36 physical component summary score. Mental and physical component summary scores were measured on continuous scales between 0 and 100. In addition, data was collected on patients demographics (gender, age, indigenous status, education, marital status) and some process of care variables (time from onset of symptoms to admission to an acute stroke unit, temperature, length of stay). The research question of interest was to see if the treatment group was significantly different from the control group in the four outcome variables. The five models that were fitted in their publication \citep{middleton2011implementation} were three logistic regressions with random effects, with modified Rankin Scale as response (cut off at 2) and Barthel index as response (cut off at 60 and 95 respectively); and two linear regressions with random effects, with mental and physical component summary scores as responses. The predictor variables included period (before and after), intervention and the interaction between period and intervention. From the significance level of the interaction term one could tell if the pre-post change in the intervention group was different from the control group. For more details of the QASC study we refer to \cite{middleton2011implementation}.

For our simulation study, to incorporate the acute stroke unit effect, we add random effects to the main model (\ref{main model}). The DPM of a Gaussian copula model with random effects on the latent space is:

\begin{equation}\label{main model with re}
\begin{aligned}
&\mathbf{z_{ic}}|\bm{\beta},\Sigma,\mathbf{b}, w=\mathbf{b_c}+\sum_{h=1}^{\infty} w_h N_{p+Q}(\mathbf{z_{i}}|(\mathbf{0},\bm{\beta_h}),\Sigma_h), \\
&\mathbf{b_c} \sim N(0,\Psi),~c=1,...,C
\end{aligned}
\end{equation}

where the random effects $\mathbf{b_c}$ are vectors of length $p+Q$ and they follow a multivariate normal distribution with mean 0 and covariance $\Psi$. The semi-conjugate prior on $\Psi$ is $Inv-Wishart~(\nu_{\Psi},\Lambda_{\Psi})$. The computational algorithms given in section 3.4 need to be adjusted accordingly and details are provided in the supplementary material. 

For the simulation design, we took the subset of the complete cases from the original data set (75.34\%) as the `true' data, and sub-sampled 300 patient from it 100 times. Then we introduced missingness that mimicked the missing data pattern in the original data set. For a detailed description, we refer to section 4.3 in \cite {wang2017copula}. Using the same assessment criteria as simulation 2, we compared the imputation accuracy and the ability to recover the parameters in some models of interest. Because we added random effects to capture the stroke unit difference, for the FCS and single copula approaches, we added random effects accordingly. For the FCS with random effects we used the `lme4' package \citep{bates2014fitting} in R, and for the single copula and infinite mixture of copulas approaches,  We wrote our own codes which is provided in the supplementary material.  
 
Table 5 and Table 6 summarises the average imputation accuracy and estimation of the parameters in the four models of interest across the 100 data sets. Overall, the copula based methods imputed more accurately than the FCS. Furthermore, our proposed DPM copula model performed better for most of the continuous variables and this may because the skewed or multimodal distributions were difficult to capture by a single copula model. Comparing the squared bias and the coverage of the parameter estimates in the four regression models, Table 6 shows that all four competing methods had reasonably good coverage rate . Although some parameters suffered from under coverage by all the methods, for example, $\beta_1$ in the model for mental health score, DPM seemed to have a coverage rate closer to the nominal 95\%. Moreover, the mean of the squared bias was almost always the smallest for the DPM copula model and the largest for the complete case method.

\begin{table}[]\label{sim3.acc}
	\centering
	\caption{Imputation accuracy in the QASC with randomly deleted records, measured by the average Euclidean distance between the imputed values and the true values for the first nine variables and misclassification rate for the last two variables, by FCS, single Gaussian copula and DPM of Gaussian copulas.}
	\label{my-label}
	\begin{tabular}{llll}
		\hline
		Variable                 & FCS    & Single copula & DPM copula \\\hline
		Time taken to hospital & 224.46  & 135.22        & 124.6      \\
		Education                 & 3.79   & 2.69          & 2.72       \\
		Age                    & 249.43 & 254.84        & 243.55     \\
		Modified Rankin Scale     & 4.16   & 2.72          & 2.49       \\
		Bartell Index           & 654.67 & 443.1        & 459.79     \\
		Physical health score   & 186.11 & 161.42        & 160.82     \\
		Mental health score     & 344.57 & 241.71        & 242.54     \\
		Length of stay        &196.01 & 158.15        & 130.32     \\
		Mean temperature          & 0.19   & 0.13          & 0.12       \\
		Marital status            & 0.55   & 0.5           & 0.52       \\
		ATSI                     & 0.02   & 0.024         & 0.021\\ \hline     
	\end{tabular}
\end{table}

\begin{table}[]\label{sim3.para}
	\centering
	\caption{A comparison of squared bias of the coefficients estimates and the corresponding 95\% coverage of the five models of interest in the QASC data set under four treatments of missing data - complete case analysis, FCS, single Gaussian copula and DPM Gaussian copulas.}
	\label{my-label}
		\scalebox{0.78}{
	\begin{tabular}{|l|lllllllll}
		\hline
		\multicolumn{2}{|l|}{\multirow{2}{*}{}} & \multicolumn{2}{l}{CC}                                   &                                  \multicolumn{2}{l}{FCS}                                & \multicolumn{2}{l}{single copula}                        & \multicolumn{2}{l|}{DPM copula}                           \\ \cline{3-10} 
		\multicolumn{2}{|l}{}                  & \multicolumn{1}{|l}{bias} & \multicolumn{1}{l}{coverage} & bias & \multicolumn{1}{l}{coverage} & bias & \multicolumn{1}{l}{coverage} & bias& \multicolumn{1}{l|}{coverage} \\ \hline
		\multirow{4}{*}{Modified Rankin Scale 2}     & \multicolumn{1}{l|}{$\beta_1$}     & 0.029                     & 91                                                   & 0.01                      & 92                            & 0.007                     & 92                            & 0.006                     & \multicolumn{1}{l|}{94}                            \\
		& \multicolumn{1}{l|}{$\beta_2$}     & 0.049                     & 94                                                  & 0.019                     & 96                            & 0.016                     & 97                            & 0.014                     & \multicolumn{1}{l|}{97}                            \\
		& \multicolumn{1}{l|}{$\beta_3$}     & 0.022                     & 93                            &  0.014                     & 100                           & 0.013                     & 99                            & 0.009                     & \multicolumn{1}{l|}{97}                            \\
		& \multicolumn{1}{l|}{$\beta_4$}     & 0.043                     & 90                            & 0.032                     & 90                            & 0.024                     & 92                            & 0.03                      & \multicolumn{1}{l|}{90}                            \\ \cline{1-1}\hline
		\multirow{4}{*}{Bartell Index 90}     & \multicolumn{1}{l|}{$\beta_1$}     & 1.228                     & 74                           &  0.16                      & 78                            & 0.123                     & 82                            & 0.121                     & \multicolumn{1}{l|}{82}                            \\
		& \multicolumn{1}{l|}{$\beta_2$}     & 3.252                     & 94                            & 1.854                     & 95                            & 1.422                     & 96                            & 1.604                     & \multicolumn{1}{l|}{94}                            \\
		& \multicolumn{1}{l|}{$\beta_3$}     & 0.263                    & 85                            &  0.162                     & 79                           & 0.175                     & 83                            & 0.16                     & \multicolumn{1}{l|}{89}                            \\
		& \multicolumn{1}{l|}{$\beta_4$}     & 2.953                     & 83                            &  1.998                     & 94                            & 1.599                    & 94                            & 1.574                      & \multicolumn{1}{l|}{93}                            \\ \cline{1-1}\hline
		\multirow{4}{*}{Bartell Index 60}     & \multicolumn{1}{l|}{$\beta_1$}     & 0.051                     & 87                            & 0.035                     & 88                            & 0.021                     & 90                            & 0.024                     & \multicolumn{1}{l|}{91}                            \\
		& \multicolumn{1}{l|}{$\beta_2$}     & 0.028                     & 97                            & 0.024                     & 95                            & 0.026                     & 95                            & 0.026                     & \multicolumn{1}{l|}{93}                            \\
		& \multicolumn{1}{l|}{$\beta_3$}     & 0.043                     & 92                            &  0.03                      & 90                            & 0.018                     & 93                            & 0.021                     & \multicolumn{1}{l|}{90}                            \\
		& \multicolumn{1}{l|}{$\beta_4$}     & 0.083                     & 93                            &  0.048                     & 97                            & 0.055                     & 98                            & 0.045                     & \multicolumn{1}{l|}{97}                            \\ \cline{1-1}\hline
		\multirow{4}{*}{Mental health score}     & \multicolumn{1}{l|}{$\beta_1$}     & 1.791                     & 86                                                     & 0.602                     & 85                            & 0.406                     & 87                            & 0.397                     &\multicolumn{1}{l|}{90}                            \\
		& \multicolumn{1}{l|}{$\beta_2$}     & 2.101                     & 98                            &  0.952                     & 97                            & 0.968                     & 99                            & 0.9                       & \multicolumn{1}{l|}{98}                            \\
		& \multicolumn{1}{l|}{$\beta_3$}     & 0.951                     & 84                            & 0.703                     & 88                            & 0.641                     & 89                            & 0.637                     & \multicolumn{1}{l|}{91}                            \\
		& \multicolumn{1}{l|}{$\beta_4$}     & 1.338                     & 96                            &  1.285                     & 100                           & 1.272                     & 100                           & 1.293                     & \multicolumn{1}{l|}{100}                           \\ \cline{1-1}\hline
		\multirow{4}{*}{Physical health score}     & \multicolumn{1}{l|}{$\beta_1$}     & 0.695                     & 85                                           & 0.309                     & 84                            & 0.197                     & 89                            & 0.19                      & \multicolumn{1}{l|}{92}                            \\
		& \multicolumn{1}{l|}{$\beta_2$}     & 0.871                     & 95                            &  0.466                     & 97                            & 0.39                      & 98                            & 0.302                     & \multicolumn{1}{l|}{97}                            \\
		& \multicolumn{1}{l|}{$\beta_3$}     & 0.684                     & 89                            &  0.391                     & 91                            & 0.331                     & 92                            & 0.348                     & \multicolumn{1}{l|}{90}                            \\
		& \multicolumn{1}{l|}{$\beta_4$}     & 0.928                     & 87                            &  0.964                     & 87                            & 0.815                     & 90                            & 0.803                     & \multicolumn{1}{l|}{88}                            \\ \cline{1-1}\hline
	\end{tabular}}
\end{table}

\section{Discussion}
In this paper, we proposed an imputation model using an infinite mixture of Gaussian copulas induced by a DPM. Multivariate data of mixed type was modeled through a joint distribution in the latent space. Compared with finite mixture model where the number of components has to be determined on an ad-hoc basis or based on some model selection criteria, our proposed nonparameteric model achieves semi-automatic order selection. However, sampling from an infinite dimensional parameter space is a daunting task, especially when the target distribution is potentially multimodal. Building upon two existing sampling methods - slice sampling and prior parallel sampling, we proposed an algorithm specific to sampling from a DPM. We showed through three simulation studies that our proposed model achieved better overall goodness of fit and was able to capture better tail dependencies compared with its single component counterpart. Simulations 2 and 3 further showed that when the data set involved missingness at random, and when the missing percentage is not negligible, our proposed model achieved better imputation accuracy and recovery of parameters estimates, especially when the variable with missing value was continuous. 

By using the prior parallel tempering algorithm, we introduced label switching deliberately. This action aids mixing in the MCMC, however, it confuses the labeling of parameters in each component. To make things more complicated, the number of mixture components varies in different iterations. The existing literature has not solved the label switching problem thoroughly, and this is one of the limitations of using Bayesian mixture models because we are not able to uniquely identify the component-specific parameters. Nevertheless, the goal of this research is not an unsupervised clustering, but rather to model the joint distribution as accurately as possible regardless of the marginal distribution of each parameter. 

Though having achieved good empirical performance, there are some issues that lack theoretical guidelines.The first issue is the choice of the kernel in the mixture model. In our first simulation, we compared the performance of using a Gaussian kernel and a t kernel, and we showed that while the single component Gaussian distribution underperformed the single component t distribution, there were no strong differences when using infinite mixtures. Admittedly, using a t distribution should achieve better robustness, however the computational burden will increase as there will no longer be a closed form solution and no natural link to the probit model for nominal variables. Also it is worthwhile exploring other classes of copula functions, for example Archimedean family of copulas. Secondly, the choice of the total mass parameter in the DP require further investigation. This parameter controls the concentration of the derived distribution, in which a lower value induces fewer clusters and a higher value induces more clusters. In our simulation studies, we arbitrarily selected a sequence of the total mass parameters, to ensure that smaller parameters could empty out redundant clusters and switchings are adequate. Not much is known about the theoretical properties of the choice of the total mass parameter and the implications on the inferential results, and this is a topic of future research.

\end{document}